\documentclass[a4paper,12pt]{article}

\usepackage{cmap}					\usepackage[T2A]{fontenc}			\usepackage[utf8]{inputenc}			\usepackage{amsmath,amsfonts,amssymb,amsthm,mathtools}

\usepackage{graphicx}  \graphicspath{{pic/}}    \usepackage{wrapfig}

\usepackage{geometry} 
\usepackage{setspace} 
\usepackage{lastpage} 
\usepackage{soul} \usepackage{epigraph}
\usepackage{hyperref}
\usepackage[usenames,dvipsnames,svgnames,table,rgb]{xcolor}
\hypersetup{				unicode=true,           pdftitle={Shape instability of lipid vesicles in uniaxial elongating flow},   pdfauthor={Maxim Shishkin},      pdfsubject={Shape instability of lipid vesicles in uniaxial elongating flow},      pdfcreator={}, pdfproducer={}, pdfkeywords={vescicle, soft matter}, colorlinks=true,       	linkcolor=red,          citecolor=black,        filecolor=magenta,      urlcolor=cyan           }

\usepackage{csquotes} 
\usepackage{float} 
\usepackage{multicol} 
\usepackage{ mathrsfs }

\usepackage{physics}

\usepackage{mathtools}

\usepackage{empheq}
\usepackage{floatflt}
\usepackage[most]{tcolorbox}
\usepackage{enumerate}

\usepackage{pgf}
\usepackage{subcaption}

\usepackage[backend=biber,bibencoding=utf8,sorting=none,maxcitenames=2,style=phys, articletitle=true, biblabel=brackets, chaptertitle=false, pageranges=false]{biblatex}
\usepackage{authblk}
\usepackage{orcidlink}
\usepackage{placeins}

\title{Spectrally Accurate Simulation of Axisymmetric Vesicle Dynamics}

\author[1,2]{M.A. Shishkin\orcidlink{0000-0002-1243-5285}}
\affil[1]{Landau Institute for Theoretical Physics of the RAS, Moscow Region, Chernogolovka 142432, Russia}
\affil[2]{HSE University, Moscow 101000, Russia}

\date{\today}

\begin{document} 

		\maketitle
		\begin{abstract}
We present a meshless numerical method for simulating the dynamics of axisymmetric vesicles in a viscous medium. Key innovations include: (1) adaptive reparameterization based on local length scales, reducing the number of required harmonics; (2) gauge dynamics for maintaining optimal parameterization; (3) error control near the symmetry axis; and (4) spectrally accurate quadrature schemes for singular integrals. The method achieves high accuracy and computational efficiency for simulating lipid bilayer dynamics and related problems in soft matter physics.
		\end{abstract}

\section{Introduction}
This note briefly outlines an approach for meshless numerical simulation of the dynamics of smooth axisymmetric surfaces, taking closed lipid bilayers---vesicles---as a representative example.
The basic step is the use of Green's functions of the linear viscous surrounding medium, which makes it possible to reduce the problem to one dimension.
Roughly speaking, we extend and improve the approach described in \cite{Veerapaneni2009} in the following aspects:
\begin{itemize}
	\item Construction of a simple reparameterization procedure based on a local characteristic length scale, which makes it possible to use a smaller number of harmonics for the same accuracy
	\item Construction of a gauge dynamics: a method for selecting the tangential velocity from the normal one so that the parameterization remains optimal
	\item Development of a scheme for preventing parametric growth of the numerical error in the computation of physical fields such as the Helfrich force near the symmetry axis
	\item Construction of a spectrally accurate quadrature scheme for evaluating integrals with singularities (based on an analytically derived singular decomposition of the axial Green's function)
\end{itemize}

\section{Parameterization of a smooth axisymmetric surface}
\begin{wrapfigure}[8]{l}{0.3\textwidth}
	\begin{center}
		\vspace{-7ex}
		\includegraphics[width=0.3\textwidth]{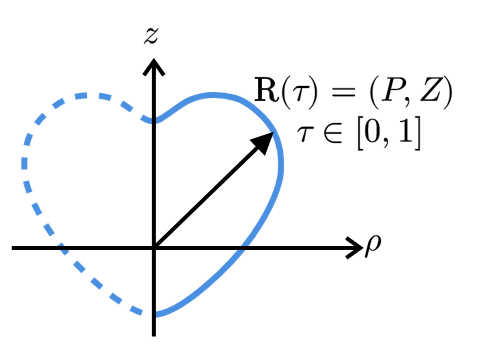}
	\end{center}
\end{wrapfigure}
An axisymmetric surface is described by a curve $\gamma$ in the plane of cylindrical coordinates $\displaystyle ( \rho, z)$, which can be parameterized by the path $\displaystyle \mathbf{R}( \tau ) \ =\ ( P( \tau ) , Z( \tau ))$.

Note that there are many parameterizations, determined by the local traversal speed along the curve: $V={dl}/{d\tau}=\sqrt{(\partial_\tau P)^2+(\partial_\tau Z)^2}$. In other words, any monotone change of the clock $\tau=\tau(\tau')$ leads to a different parameterization of the curve $\gamma$: $\displaystyle \mathbf{R} '( \tau ') =\mathbf{R}( \tau ( \tau '))$.

There is a smoothness condition for physical fields on the axis: $\displaystyle \partial _{\rho } \Psi \left( \rho \rightarrow 0\right) =0$, which is convenient to enforce automatically by choosing a parameterization for which the evenness condition holds:
\begin{equation}\label{eq:bound_gauge}
\partial _{\tau }^{2j+1} \Psi \left( \rho \rightarrow 0\right) =0,\ \partial _{\tau }^{2j} P\left( \rho \rightarrow 0\right) =0,
\end{equation}
which is assumed by default in what follows.

The simplest way to satisfy the boundary conditions~\ref{eq:bound_gauge} is to choose an appropriate functional basis; for this, we may conceptually reflect the curve across the $\displaystyle z$ axis, obtaining a smooth closed curve that is a diffeomorphism of the circle, i.e., all smooth physical fields are smooth periodic fields on the interval $\displaystyle \tau \in [ 0,2]$.
\subsection{Fourier basis}
It appears quite natural to use, as a finite-dimensional approximation of the membrane shape (the curve $\gamma$), a finite initial segment of the Fourier series \cite{Veerapaneni2009}:
\begin{equation}
    \mathbf{R}_N(\tau)= \begin{pmatrix}
P_N( \tau ) =\sum_{k=1}^{k=N} p_{k} \varphi _{\rho k}\\
Z_N( \tau ) =\sum_{k=0}^{k=N} z_{k} \varphi _{k}
\end{pmatrix}.
\label{eq:finite_series}
\end{equation}

\begin{wrapfigure}[9]{l}{0.2\textwidth}
	\begin{center}
		\vspace{-3ex}
		\includegraphics[width=0.2\textwidth]{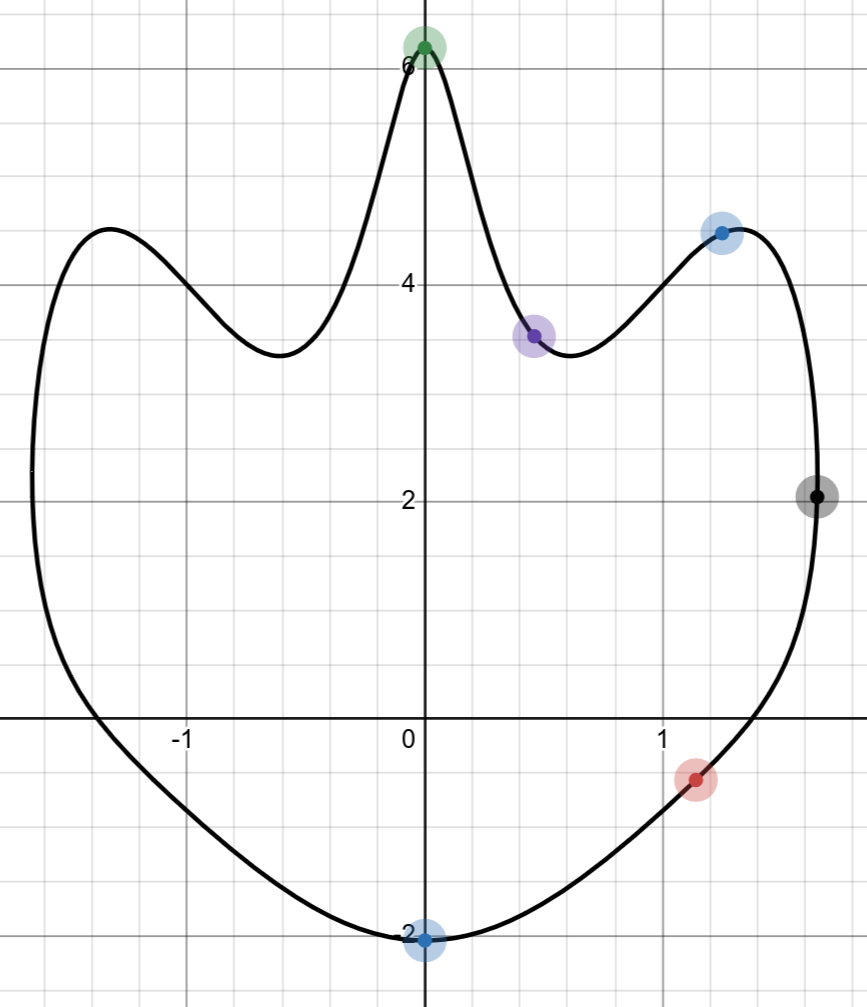}
	\end{center}
\end{wrapfigure}
We call the best finite-dimensional approximation of a \textit{given} parameterization $\mathbf{R}(\tau)$ the one that minimizes the $L_2(d\tau)$ norm, i.e., the one that is simply the truncated Fourier series of the functions in $\mathbf{R}(\tau)$.
Recall that the finite-dimensional approximation $\mathbf{R}_N(\tau)$ obtained in this way is also the best one for any $L_2$ metric on the derivatives.
It should be noted that, in practice, one often has to specify nontrivial membrane shapes from an image---either from an experiment or from a sketch.
Therefore, for practical use, a tool for smooth interpolation of the surface shape from a set of $N+2$ points in the plane $\{(P_j, Z_j)\}$ is extremely useful:
Such a problem is solved by a modification of the Kotelnikov sampling formula for our case; see \ref{pril_1}.
\subsection{Gauge}
Again, in principle one may pose explicitly the question of an optimal gauge, for example, for a prescribed number of Fourier harmonics $N$, obtained by a finite segment of the Fourier series, and an invariant measure $\mathcal{P}(\mathbf{R},\mathbf{R}_N)$ between the true shape and the shape obtained with a finite Fourier series.
\begin{equation}
    V_{\mathrm{opt}}(\tau): \mathcal{P}(\mathbf{R},\mathbf{R}_N)\to \min
\end{equation}

However, the problem of finding such a gauge remains very difficult and poorly conditioned, so we would like to be able to construct empirically, although not the best, a \textit{good} gauge.

The choice of the speed $V(\tau)=\frac{dl}{d\tau}$ corresponds to choosing the density of points $\{\mathbf{R}_j=\mathbf{R}(\tau_j)\}$ on the curve in physical space for a uniform density of points in $\tau$-space $\{\tau_j\propto j\}$.
At the level of physical intuition, we understand that in “narrow” regions the point density should be increased.
Therefore, it is sufficient to devise a rule for estimating a local length scale $a_\mathrm{loc}$ and to use a gauge with the qualitative principle: the smaller the local scale, the larger the point density:
\begin{equation}
    V(\tau)\propto a_\mathrm{loc}(\tau).
\end{equation}

This leads to the problem of defining the local scale $a_{loc}$.
The sources of information about the scale can naturally be divided into groups:
\begin{itemize}
    \item those associated with the planar curve $\gamma $, such as the curvature $H_1 = \frac{P''Z'-P'Z''}{V^{3}}$ together with its derivative ${d H_1}/dl$;
    \item those associated with the presence of the symmetry axis, i.e., $P$ and the second curvature $H_2 = \frac{-Z'}{P V}$;
    \item the overall characteristic size $R_s$;
    \item scales associated with the structure of the external forcing
\end{itemize}

Based on estimates of the various scales $a_i$, one needs to assemble a final estimate of the local scale, for example using a weighted Kolmogorov mean.
Note that, by construction, $a_{loc}$ should not vary on a scale smaller than $a_{loc}$.

One of the simplest and most natural choices is the weighted power mean (with $\alpha<0$, since $a_i\to\infty$ should not contribute to the estimate)
\begin{equation}
    a_{loc}(\tau)=\left(\sum_i w_i a_i^\alpha /\sum_i w_i \right)^{1/\alpha}.
\end{equation}
The simplest scheme assumes the use of ${\alpha=-2}$ with the scales $a_i=\{1/|H1|, 1/|H2|, R_s\}$, while the choice of weights $w_i$ depends on the specific problem.

After specifying a rule for estimating the local scale and the corresponding desired traversal speed $V(\tau)$, the problem of re-gauging naturally arises---replacing the current parameterization with a new one $\mathbf{R}(\tau)\to \mathbf{R'}(\tau)$ such that the speed $V'\propto a_{loc}$; this is addressed in \ref{app:regauge}.
\section{Gauge dynamics}
In addition to solving static problems on a prescribed surface shape, one faces the problem of simulating shape dynamics:
$$
\frac{\partial \mathbf{\mathbf{R}}}{\partial t} = \mathbf{F}[t, \mathbf{R}],
$$
while, from the viewpoint of shape dynamics, only the normal component of the velocity is important: $v_n = \mathbf{l}\cdot \mathbf{F}$, where $\mathbf{l}$ is the local normal vector to the surface.
The tangential component, by contrast, can be chosen in any convenient way.
Since, for numerical computations, it is convenient to use what provides the best accuracy, it is reasonable to exploit this freedom; the most natural way is to choose the tangential velocity so as to preserve the gauge, i.e., so that if at the initial time the representation is such that
$V(\tau, t_0)\propto a_{loc}(\tau)[\mathbf{R}_0]$, the same relation will hold at any subsequent time.\footnote{Note that if one does not control the tangential dynamics, rarefactions in the point density may well appear in newly formed “narrow” regions. In principle, one may run the dynamics for a limited time interval, until the representation becomes significantly degraded, and then perform the re-gauging procedure; however, this is not very convenient, since it requires manual intervention.}

Note that, in mesh-based methods, this approach corresponds to the broad topic of Mesh-Control and Regridding\cite{Pozrikidis2001}.

\subsection{Constant speed}
\href{https://en.wikipedia.org/wiki/Differentiable_curve}{Natural parametrization}

One can construct a functional ${\displaystyle \mathbf{v}_{g} [\mathbf{v}_{n} ]}$ that compensates for the violation of the gauge in the tangent bundle.

Let there be an infinitesimal change:
\begin{equation*}
\delta \mathbf{R} \ =\ \begin{pmatrix}
\delta P\\
\delta Z
\end{pmatrix} \ =\ \mathbf{l} \ \delta R_{l} \ +\ \mathbf{e}_{\tau } \ \delta R_{\tau } ,
\end{equation*}
Then the speed field ${ {dl}/{d\tau } =V}$ transforms as follows:
\begin{equation*}
\delta V\ =\ \mathbf{e}_{\tau } \partial _{\tau } \delta \mathbf{R} \ =(H_{1} V)\cdot \delta R_{n} +\partial _{\tau } \delta R_{\tau }
\end{equation*}
If we want to maintain a uniform speed, ${\displaystyle \delta R_{\tau }}$ must be determined from the requirement that the variation of the speed be equal to a constant\footnote{The total curve length, in dynamics, is of course not conserved, unlike the area.}:
\begin{gather*}
\delta V=C\Longrightarrow 
\partial _{\tau } \delta R_{\tau } =-(H_{1} v)\cdot \delta R_{n} +C,
\end{gather*}
whence ${\displaystyle R_{\tau }}$ can be found explicitly in terms of the Fourier components:
\begin{equation*}
\delta R_{\tau |k} =-\frac{\| (H_{1} v)\cdot \delta R_{n} \| _{k}}{\partial _{\tau |k}} ,\ k >0
\end{equation*}
The final expression is:
\begin{equation*}
\delta \mathbf{R}_{corr} \ =\ \mathbf{l} \ \delta R_{n} \ +\ \mathbf{e}_{\tau } \ \left\Vert -\frac{\| (H_{1} v)\cdot \delta R_{n} \| _{k}}{\partial _{\tau |k}}\right\Vert _{\tau } .
\end{equation*}
In a similar manner, expressions can be obtained for preserving an arbitrary gauge.
\subsection{Determining the tangential dynamics }

We want, to linear order in $\displaystyle \delta R$, to solve the equation:
\begin{gather*}
V[\mathbf{R}]( \tau ) =Ca_{loc}[\mathbf{R}] \ ( \tau )\\
\delta \mathbf{R} \ =\ \begin{pmatrix}
\delta P\\
\delta Z
\end{pmatrix} \ =\ \mathbf{l} \ \delta R_{l} \ +\ \mathbf{e}_{\tau } \ \delta R_{\tau }
\end{gather*}
Take the variation:
\begin{gather*}
\left(\frac{\delta V}{\delta \mathbf{R}} ,\mathbf{l} \delta R_{n}\right)( \tau ) +\left(\frac{\delta V}{\delta \mathbf{R}} ,\mathbf{e}_{\tau } \delta R_{\tau }\right)( \tau ) =
\delta C\ a_{loc}[\mathbf{R}] \ ( \tau ) +C\left(\frac{\delta a_{loc}}{\delta \mathbf{R}} ,\mathbf{l} \delta R_{n}\right)( \tau ) +\left(\frac{\delta a_{loc}}{\delta \mathbf{R}} ,\mathbf{e}_{\tau } \delta R_{\tau }\right)( \tau )
\end{gather*}

\begin{gather}
\left(\frac{\delta V}{\delta \mathbf{R}} -C\frac{\delta a_{loc}}{\delta \mathbf{R}} ,\mathbf{e}_{\tau }\textcolor[rgb]{0.07,0.2,0.94}{\delta R_{\tau }}\right) -\textcolor[rgb]{0.07,0.2,0.94}{\delta C} \ a_{loc}[\mathbf{R}] =
\textcolor[rgb]{0.82,0.01,0.11}{\left( C\frac{\delta a_{loc}}{\delta \mathbf{R}} -\frac{\delta V}{\delta \mathbf{R}} ,\mathbf{l} \ \delta R_{n}\right)}
\end{gather}
Note that if $\displaystyle a_{loc}$ is a scalar field, then its variation under a tangential displacement is associated only with advection of its value, so we can write the left-hand side immediately:
\begin{equation}
\left[ \partial _{\tau }\textcolor[rgb]{0.07,0.2,0.94}{\ } -\frac{\partial _{\tau } a_{loc}}{a_{loc}}\right]\textcolor[rgb]{0.07,0.2,0.94}{\delta R}\textcolor[rgb]{0.07,0.2,0.94}{_{\tau }} =\textcolor[rgb]{0.82,0.01,0.11}{\left( C\frac{\delta a_{loc}}{\delta \mathbf{R}} -\frac{\delta V}{\delta \mathbf{R}} ,\mathbf{l} \ \delta R_{n}\right)}\textcolor[rgb]{0.07,0.2,0.94}{+\delta C} a_{loc}
\end{equation}
Despite the somewhat intimidating appearance of this equation, its solution can be written explicitly:
\begin{gather}
\textcolor[rgb]{0.07,0.2,0.94}{\delta R_{\tau }} =a_{loc} \cdot \textcolor[rgb]{0.07,0.2,0.94}{f} \Longrightarrow 
\partial _{\tau }\textcolor[rgb]{0.07,0.2,0.94}{f} =\textcolor[rgb]{0.82,0.01,0.11}{\frac{1}{a_{loc}}\left( C\frac{\delta a_{loc}}{\delta \mathbf{R}} -\frac{\delta V}{\delta \mathbf{R}} ,\mathbf{l} \ \delta R_{n}\right)}\textcolor[rgb]{0.07,0.2,0.94}{+\delta C} \ 
\end{gather}
From this, by analogy with the previous item, we easily obtain the final result.
\begin{equation*}
\boxed{\textcolor[rgb]{0.07,0.2,0.94}{f}_{k} =\frac{\| \textcolor[rgb]{0.82,0.01,0.11}{\frac{1}{a_{loc}}\left( C\frac{\delta a_{loc}}{\delta \mathbf{R}} -\frac{\delta V}{\delta \mathbf{R}} ,\mathbf{l} \ \delta R_{n}\right)} \| _{k}}{\partial _{\tau |k}} ,\ k >0}
\end{equation*}
Note that, in addition to such a rigid correction of the dynamics, one can also use a softer one---a $\mathbf{v}_\tau$ for which the equilibrium relation is $V(\tau)\propto a_\mathrm{loc}$.
\section{Viscous dynamics of lipid bilayers}
We illustrate the application of the scheme by simulating the dynamics of lipid bilayers,
which, from the hydrodynamic viewpoint, are two-dimensional films immersed in a viscous fluid, with free energies localized on them and, as a consequence, stresses.
\subsection{Dissipative equations}
In the hydrodynamic limit, we treat lipid bilayers as infinitely thin liquid films of various shapes immersed in a viscous fluid. This approach assumes characteristic shape variations on a length scale much larger than the bilayer thickness ($\displaystyle h\sim 10^{-7}\text{cm}$), which holds for typical vesicle experiments ($R\sim 10 \,\mu\text{m}$). We study membrane dynamics, i.e., the evolution of its shape in time.
\subsubsection{Membrane stresses}
The membrane state is specified by its shape and the surface number density of lipid molecules $\displaystyle n$. The membrane shape is associated with the Helfrich bending energy\cite{Helfrich1973}:
\begin{equation}
\mathcal{F} \ =\ \int _{S}\frac{\kappa }{2}\left(\frac{1}{R_{1}} +\frac{1}{R_{2}}\right)^{2} ,
\end{equation}
where the integration is over the membrane surface, with the Helfrich modulus $\displaystyle \kappa \sim 20\ k_{B} T$\footnote{The values of the energy moduli are given for convenience in units corresponding to room temperature.} and principal radii of curvature $\displaystyle R_{1} ,R_{2}$. \footnote{In addition to the expression given, at leading hydrodynamic order one may also write another invariant contribution to the surface free-energy density, proportional to the Gaussian curvature $\displaystyle K\varpropto \frac{1}{R_{1} R_{2}}$, which we omit since, by the Gauss--Bonnet theorem, such a contribution to the energy is a topological invariant and does not contribute to the stress tensor of a homogeneous membrane.}

To leading order in the deviation of the surface density $\displaystyle n$ from its equilibrium value $\displaystyle n_{0}$, the surface energy density associated with membrane stretching is:
\begin{equation}
\varepsilon =\frac{K_{n}}{2}\left(\frac{n}{n_{0}} -1\right)^{2} ,
\end{equation}
where $\displaystyle K_{n} \sim 60\ \frac{k_{B} T}{\mathrm{nm}^{2}}$ is the membrane stretching modulus. Then the surface tension is
\begin{equation}
\sigma =\varepsilon -n\partial _{n} \varepsilon =-K_{n}\left(\frac{n}{n_{0}} -1\right) .
\end{equation}
If the membrane is out of equilibrium, stresses localized on it arise; these induce fluid motion around it, which in turn advects the membrane.

Surface forces naturally split into the Helfrich force and the force associated with surface tension\footnote{A detailed derivation of the stress tensor in the nondissipative approximation is given in Appendix~\ref{pril_1}.}:
\begin{gather}
\mathbf{f}_{\kappa } =\kappa \ \mathbf{l}\left(\frac{H^{3}}{2} -2HK+\Delta ^{\perp } H\right) , \label{eq:Helfrih_force} \\
\mathbf{f}_{\sigma } =\nabla ^{\perp } \sigma -\mathbf{l} \ \sigma H,
\end{gather}
where $\displaystyle H=\frac{1}{R_{1}} +\frac{1}{R_{2}}$ is the mean curvature, $\displaystyle K=\frac{1}{R_{1} R_{2}}$ is the Gaussian curvature, $\displaystyle \mathbf{l}$ is the normal vector to the bilayer surface, and $\displaystyle \nabla ^{\perp }$ and $\displaystyle \Delta ^{\perp }$ are the gradient and Beltrami--Laplace operators on the surface manifold of the membrane.

\subsubsection{Quasistatics of the surrounding fluid}
A lipid bilayer is a rather soft object; therefore, the fluid motion around it is typically slow and at low velocity, and the fluid velocity in the bulk satisfies the Stokes equation (in the absence of external forcing on the fluid):
\begin{equation*}
-\nabla p\ +\ \eta \Delta \mathbf{v}=0 ,
\end{equation*}
where $\displaystyle \mathbf{v}$ is the flow velocity,where $\displaystyle p$ is the pressure, and $\displaystyle \eta $ is the dynamic viscosity coefficient. Due to the assumed slowness of the flow, it can be considered incompressible $\displaystyle ( \nabla \mathbf{v}) =0$, which determines the pressure $\displaystyle p$. 

\subsubsection{Coupling between membrane dynamics and the surrounding fluid}
The last equation describes the dynamics of the surface density $\displaystyle n$:
\begin{equation*}
\partial _{t} n\ =\ -\mathbf{v} \nabla n\ -n\nabla ^{\perp }\mathbf{v} \Longrightarrow \partial _{t} \sigma \ =\ -\mathbf{v} \nabla \sigma \ -\sigma \nabla ^{\perp }\mathbf{v} +\textcolor[rgb]{0.82,0.01,0.11}{K}\textcolor[rgb]{0.82,0.01,0.11}{_{n}} \nabla ^{\perp }\mathbf{v} ,
\end{equation*}
which, due to weak compressibility, simplifies to
\begin{equation*}
\frac{d}{dt} \sigma =\partial _{t} \sigma \ +\mathbf{v} \nabla \sigma \ =\textcolor[rgb]{0.82,0.01,0.11}{K}\textcolor[rgb]{0.82,0.01,0.11}{_{n}} \nabla ^{\perp }\mathbf{v} ,
\end{equation*}
describing relaxation of the surface tension $\displaystyle \sigma $ over a short time $\displaystyle \varpropto \frac{1}{\textcolor[rgb]{0.82,0.01,0.11}{K}\textcolor[rgb]{0.82,0.01,0.11}{_{n}}}$ to a quasi-equilibrium value at which the surface divergence of the velocity vanishes:
\begin{equation}
\boxed{\nabla ^{\perp }\mathbf{v} =0}\footnote{Surface\ tension,\ similarly\ to\ pressure\ in\ classical\ hydrodynamics,\ is\ not\ a\ dynamical\ variable,\ enforcing\ surface\ divergence-freeness\ of\ the\ velocity\ field.} .
\end{equation}
Thus, we neglect the surface-tension dynamics in the parameter $\displaystyle \frac{\kappa R^{2}}{\textcolor[rgb]{0.82,0.01,0.11}{K}\textcolor[rgb]{0.82,0.01,0.11}{_{n}}} \ll 1$. Since, by the Stokes equation, the velocity is linear in the boundary conditions and hence in the surface tension, the divergence-free condition on the velocity is a linear equation for $\displaystyle \sigma $.\footnote{Surface tension induces a contribution to the membrane stress tensor such that the surface divergence of the velocity induced by it cancels the surface divergence of the flow generated by all other sources.}

\begin{figure} \centering
    \includegraphics[width=0.5\linewidth]{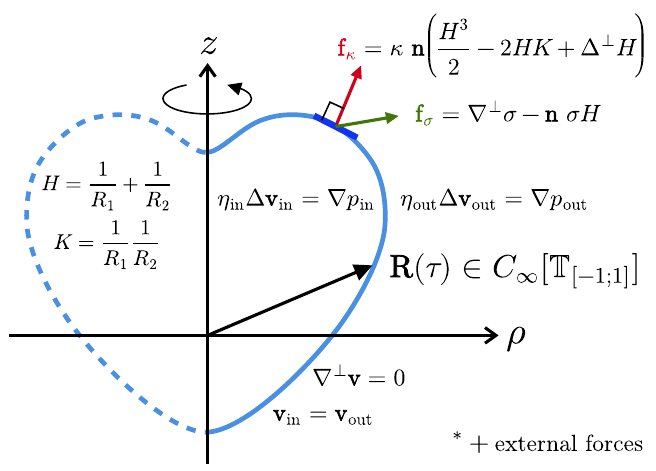}
    \caption{Schematic representation of the equations governing membrane dynamics.}
    \label{fig:scheme_equation}
\end{figure}

Thus, we arrive at a closed description of membrane dynamics in terms of its shape: the surface tension and the surrounding-fluid flow are determined quasi-statically by the current surface shape. The shape changes due to the no-slip and impermeability conditions at the membrane; literally, it is advected by the flow $\displaystyle \mathbf{v}_{\mathrm{mem}} =\mathbf{v}$. A schematic description of the dynamics is given in Fig.~\ref{fig:scheme_equation}.

\subsection{Determining the bulk fluid motion}
\subsubsection{Surface-integration method}
For simplicity, in what follows we consider the case where the fluid viscosities inside and outside the membrane are equal\footnote{In the case of different viscosities, the membrane velocity can be obtained without resolving the flow in the fluid bulk by solving a boundary integral equation on the surface\cite{Pozrikidis1992}.}, so that the flow velocity $\displaystyle \mathbf{v}$ satisfies a single equation throughout all space,
\begin{equation}
\partial _{k} p-\eta \Delta v_{k} =\hat{f}_{k} ,
\end{equation}
with the right-hand side being a source localized on the membrane. This equation can be solved explicitly using the Oseen Green's function:
\begin{gather}
v_{i} (\mathbf{r} )=\int \dd V_{1} G_{ik} (\mathbf{r} -\mathbf{r_{1}} )\hat{f}_{1}^{k} , \\
G_{ik} (\mathbf{r} )=\frac{1}{8\pi \eta r}\left( \delta _{ik} +\frac{r_{i} r_{k}}{r^{2}}\right) .
\end{gather}
It is convenient to rewrite the resulting relation in terms of the surface force density:
\begin{equation}
v_{i} (\mathbf{r} )=\int \dd S_{1} G_{ik} (\mathbf{r} -\mathbf{r_{1}} )f_{1}^{k} .
\end{equation}
\subsubsection{Green's function for the axisymmetric case}

Below, we simulate processes in which the membrane shape is axisymmetric. To simplify the computations, it is necessary to carry out the integration over the azimuthal angle explicitly, yielding a reduced Green's function in cylindrical coordinates:
\begin{equation*}
v_{\mathbf{\mu }} (\rho ,z)=\int dV_{1} G_{\mu \nu } (\rho ,\rho _{1} ,z-z_{1} )f_{1}^{\nu } ,\ \text{where} \ \mu ,\nu \in [ \rho ,z]
\end{equation*}
The corresponding expressions are given in \ref{app:green_axial} and have a form equivalent to that in \cite{Pozrikidis1992}, but more convenient for numerical manipulations.

It should be noted that the Green's function has a logarithmic singularity:
\begin{equation*}
G_{\mu \nu } \sim \delta _{\mu \nu }\textcolor[rgb]{0.82,0.01,0.11}{\frac{2}{\rho }}\textcolor[rgb]{0.82,0.01,0.11}{\ln}\textcolor[rgb]{0.82,0.01,0.11}{(4}\textcolor[rgb]{0.82,0.01,0.11}{\frac{2\rho }{\delta r}}\textcolor[rgb]{0.82,0.01,0.11}{)} \ .
\end{equation*}


\subsection{Determining the surface tension}
Since the dependence of the velocity on the surface tension can be expressed via the Green's function, we obtain the equation
\begin{equation*}
\left( \nabla ^{\perp }\mathbf{v}\right) =\underbrace{\int\nolimits _{S_{1}}\left( \nabla ^{\perp }\hat{G}\right)\left[ \nabla _{1}^{\perp } -\mathbf{n}_{1} H_{1}\right]\textcolor[rgb]{0.25,0.46,0.02}{\sigma }\textcolor[rgb]{0.25,0.46,0.02}{_{1}}}_{\left( \nabla ^{\perp }\mathbf{v}_{\sigma }\right)[\textcolor[rgb]{0.25,0.46,0.02}{\sigma }]}\textcolor[rgb]{0.25,0.46,0.02}{+}\left( \nabla ^{\perp }\mathbf{v}\right)_{-\sigma } =0,
\end{equation*}
where $\displaystyle \left( \nabla ^{\perp }\mathbf{v}\right)_{-\sigma }$ denotes the surface divergence induced without taking the surface tension into account, and $\displaystyle \left( \nabla ^{\perp }\mathbf{v}_{\sigma }\right)$ is a linear operator with kernel $\displaystyle \left( \nabla ^{\perp }\hat{G}\right)\left[ \nabla _{1}^{\perp } -\mathbf{n}_{1} H_{1}\right]$.
Note that all eigenvalues of this linear operator are nonpositive, which corresponds precisely to relaxation of the surface tension toward equilibrium.

An explicit form in terms of surface forces is:
\begin{gather*}
\left( \nabla ^{\perp }\mathbf{v}\right) (\tau )\ =\ \int _{S_{1}}\mathbf{D} (\tau ,\tau _{1} )\cdot \mathbf{f} (\tau _{1} )\\
D_{\rho } \ =\ \frac{P_{\tau }^{2}}{v^{2}} \partial _{\rho } G_{\rho \rho } \ +\frac{Z_{\tau }^{2}}{v^{2}} \partial _{z} G_{z\rho } +\ \frac{1}{v^{2}} P_{\tau } Z_{\tau } (\partial _{z} G_{\rho \rho } +\partial _{\rho } G_{z\rho } )\ +\ \frac{G_{\rho \rho }}{P}\\
D_{z} \ =\ \frac{P_{\tau }^{2}}{v^{2}} \partial _{\rho } G_{\rho z} \ +\frac{Z_{\tau }^{2}}{v^{2}} \partial _{z} G_{zz} +\ \frac{1}{v^{2}} P_{\tau } Z_{\tau } (\partial _{z} G_{\rho z} +\partial _{\rho } G_{zz} )\ +\ \frac{G_{\rho z}}{P}
\end{gather*}

We will compute $\displaystyle \sigma $ approximately using a finite segment of the Fourier series,
\begin{equation}
\sigma =\sum _{n=0}^{n=N_{\sigma }} \sigma _{n} \varphi _{n} ,
\end{equation}
by minimizing the $\displaystyle L_{2}$ norm of the equation residual with respect to its coefficients:
\begin{equation}
\int \left(\left( \nabla ^{\perp }\mathbf{v}_{\sigma }\right)[\textcolor[rgb]{0.25,0.46,0.02}{\sigma }] \ +\left( \nabla ^{\perp }\mathbf{v}\right)_{-\sigma }\right)^{2} \ d\mu \rightarrow \min
\end{equation}
The choice of measure $\displaystyle \mu $ is a matter of modeling design; natural candidates are $\displaystyle dS,dl,d\tau $. In any case, for the coefficients $\{\sigma_n\}$ we obtain a linear problem:
\begin{gather}
\sum_n \mathcal{\Vert O_{\sigma }\Vert }_{kn}\sigma_n \ =-\mathbf{\Vert s_{\sigma }\Vert }_{k} ,\ \\
\mathcal{\Vert O_{\sigma }\Vert }_{kn} =\int \left( \nabla ^{\perp }\mathbf{v}_{\sigma }[ \varphi _{n}]\right)\left( \nabla ^{\perp }\mathbf{v}_{\sigma }[ \varphi _{k}]\right) d\mu ,\\
\ \mathbf{\Vert s_{\sigma }\Vert }_{k} =\int \left( \nabla ^{\perp }\mathbf{v}\right)_{-\sigma }\left( \nabla ^{\perp }\mathbf{v}_{\sigma } \varphi _{k}\right) d\mu ,
\end{gather}
which is not severely ill-conditioned due to diagonal dominance and the larger number of points at which the surface divergence is evaluated, $n_\mathrm{int}>N_\sigma$.
\subsubsection{Quadrature approximation}
\label{quad_approx}
Consider, for example, the measure $\displaystyle d\mu =d\tau $; then, due to smoothness, the optimal (Gaussian) quadrature rule for integrating symmetric fields is $\displaystyle \int fd\tau \approx \frac{\frac{1}{2} f_{0} +f_{1} +\cdots \frac{1}{2} f_{n}}{n} =c_{j} f_{j}$.

Consider the matrix with the first index in physical space, $\displaystyle M_{jk} =\left( \nabla ^{\perp }\mathbf{v}_{\sigma }[ \varphi _{k}]\right)_{j}$. Let us rescale it to account for the measure $\displaystyle d\mu $, $\displaystyle \tilde{M}_{jk} =\sqrt{c_{j}} M_{jk}$. The vector $\displaystyle \mathbf{rhs}_{j} =\left\Vert -\left( \nabla ^{\perp }\mathbf{v}\right)_{-\sigma }\right\Vert _{j}$ is rescaled similarly, $\displaystyle \widetilde{\mathbf{rhs}}_{j} =\mathbf{rhs}_{j}\sqrt{c_{j}}$; then the equation takes the form
\begin{equation}
\tilde{M}^{T}\tilde{M} \ \mathbf{\sigma } \ =\ \tilde{M}^{T}\widetilde{\mathbf{rhs}} ,
\end{equation}
so that the solution is also the solution of a linear regression problem with the rectangular matrix $\displaystyle \tilde{M}$.

(This makes it possible to avoid multiplying by the matrix $\displaystyle \tilde{M}^{T}$ by using robust LLS methods.)
\section{Evaluation of observables near the symmetry axis}
Since the azimuthal curvature $H_2$ contains an indeterminacy $ \varpropto {Z'}/{P}$, the Laplace--Beltrami operator applied to it contains terms whose evaluation in the vicinity of the symmetry axis $\tau =\varepsilon \sim 1/N_{\mathrm{int}}$ leads to a loss of accuracy. In the best case, after the Fourier transform we can expect machine relative accuracy for each derivative. 
\begin{equation*}
-v H_{2}( \varepsilon ) = \frac{Z'}{P}( \varepsilon ) \approx \frac{Z_{2} \varepsilon ( 1+\epsilon _{arh})}{P_{1} \varepsilon } =\frac{Z_{2}}{P_{1}}( 1+\epsilon _{arh}) ,
\end{equation*}
where we have omitted the argument $(\tau=0)$ in the derivatives.
The expression for the Helfrich force \eqref{eq:Helfrih_force} involves the Laplacian $\frac{1}{vP}\left( \partial _{\tau }\frac{P}{v} \partial _{\tau }\right)$ of this curvature.
Let us track how the accuracy deteriorates when evaluating this expression by direct substitution.
The first derivative in the vicinity of the axis,
\begin{gather*}
\frac{Z''P-P'Z'}{P^{2}} \approx \frac{\left( Z_{2} +\frac{Z_{4} \varepsilon ^{2}}{2}\right)\left( P_{1} \varepsilon +\frac{P_{3} \varepsilon ^{3}}{3!}\right)( 1+\epsilon _{arh}) -\left( Z_{2} \varepsilon +\frac{Z_{4} \varepsilon ^{3}}{3!}\right)\left( P_{1} +\frac{P_{3} \varepsilon ^{2}}{2}\right)\left( 1+\tilde{\epsilon }_{arh}\right)}{P_{1}^{2} \varepsilon ^{2}} \approx \\
\frac{Z_{2} P_{1} \varepsilon \epsilon _{arh} \ +\varepsilon ^{3}\frac{1}{3}( P_{1} Z_{4} -Z_{2} P_{3})}{P_{1}^{2} \varepsilon ^{2}} \approx \varepsilon \left[\frac{\frac{1}{3}( P_{1} Z_{4} -Z_{2} P_{3})}{P_{1}^{2}} +\frac{Z_{2} P_{1} \ }{P_{1}^{2}}\textcolor[rgb]{0.82,0.01,0.11}{\frac{\epsilon _{arh}}{\varepsilon }}\right]
\end{gather*}
already has a relative accuracy worse by a factor of $\displaystyle \varepsilon $.

We propose the following solution to the loss-of-accuracy problem near the axis. 
Since both quantities can be written explicitly as $\displaystyle \sin$ series:
\begin{equation*}
Z'\ =\ \sum _{k=1}^{k=N} z_{k} 2(\cos( k\pi \tau )) '=\sum _{k=1}^{k=N} z_{k}( -k\pi ) 2(\sin( k\pi \tau )) ',
\end{equation*}
the indeterminacy \ can be eliminated by explicitly canceling the sine factor within each series, using the basic identity:
\begin{equation}
\frac{\sin( 2n+1) x}{\sin x} =1+\sum _{k=1}^{k=n} 2\cos( 2kx) ,\ \frac{\sin 2nx}{\sin x} =\sum _{k=1}^{k=n} 2\cos(( 2k-1) x) .
\end{equation}

\begin{figure}[H]
	\centering
	\begin{subfigure}[b]{0.15\textwidth}
		\centering
		\includegraphics[width=\textwidth]{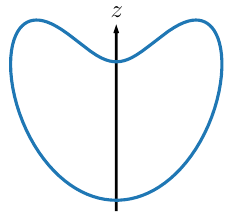}\vspace{6ex}
		\caption{}
		\label{fig:Shape_for_curvature_demo}
	\end{subfigure}\hfill
	\begin{subfigure}[b]{0.45\textwidth}
		\centering
		\includegraphics[width=\textwidth]{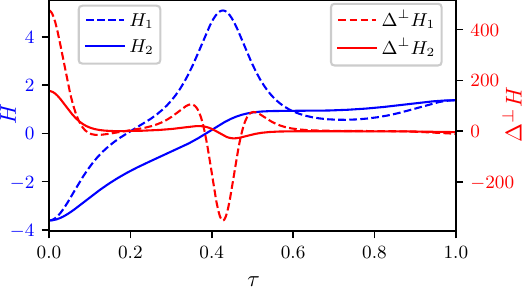}
		\caption{}
		\label{fig:Curvature_and_lapl_for_curvature_demo}
	\end{subfigure}\hfill
	\begin{subfigure}[b]{0.35\textwidth}
	\centering
	\includegraphics{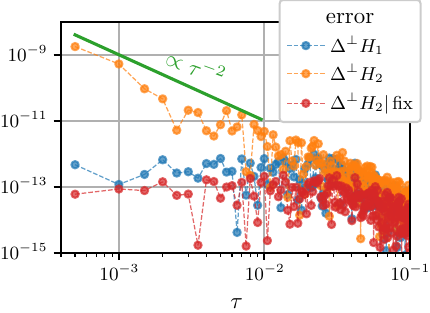}
	\caption{}
	\label{fig:error_lapl_for_curvature_demo}
	\end{subfigure}
	\caption{Demonstration of suppressing the growth of the error in computing the Laplacian of the surface curvature near the axis. For a vesicle of the shape (\subref{fig:Shape_for_curvature_demo}) parameterized by the set $(\{p_j\}, \{z_j\})=(\{0,1,-0.2\}, \{0, 0.7, -0.5\})$, panel (\subref{fig:Curvature_and_lapl_for_curvature_demo}) shows the profiles of the two principal curvatures and their Laplacians. Panel (\subref{fig:error_lapl_for_curvature_demo}) presents the absolute local errors in computing the Laplacian of the curvatures by direct differentiation and by representing $H_2$ via division of $\sin$ series by sine, which, as can be seen, eliminates the parametric increase of the error. }
	\label{fig:enh_lapl_curvature}  
\end{figure}

An example of suppressing the growth of the error when approaching the symmetry axis is shown in Fig. \ref{fig:enh_lapl_curvature}

\section{Evaluation of integrals}
To determine the induced flow on the membrane due to surface forces using the Green's function, one has to evaluate integrals with singular parts numerically. This is the most computationally demanding component, since one needs not only to approximate the resulting velocity with sufficient accuracy, but also to do so accurately enough for the dynamics to remain stable even for the high-frequency part of the parameterization; thus, fairly high accuracy is required in the case where the integrand varies on a scale of order $1/N$. Of course, the problem of integrating functions with a logarithmic singularity is well known in numerical analysis\cite{Wu2021}\cite{Hao2012}: there exist both weight-correction formulas on a uniform grid (Kapur--Rokhlin)\cite{Kapur1997} and more robust approaches employing additional nodes near the singularity (Alpert)\cite{Alpert1999}. 
In addition, general-purpose methods such as the tanh method \cite{Press2007} are also applicable in our case of a weak logarithmic singularity. In practice, however, what matters is not the mere possibility of evaluating such an integral, but the required computational time—one must compute such an integral $n_\mathrm{int}$ times—so we would like to do this as fast as possible for a given accuracy level. Below we present methods that achieve spectral accuracy while remaining computationally simple.

Let us briefly comment on the specifics of the problem. Since, for the response $v_\mu (\tau_o)$, the integrand singularity is structured as a sum of a regular part and a regular function multiplied by a logarithm, an arbitrarily accurate quadrature formula requires analytical knowledge of the regular factor multiplying the logarithm. We were able to obtain such a relation by deriving the following local branching properties for elliptic functions:
In terms of the inverse variable (for convenience, it tends to zero at the singular point)\footnote{We obtained this result directly from the coefficients of the series expansion; we would appreciate a reference to a work where this result was proven earlier. In addition, it is of interest whether expressions in terms of special functions are known for the regular contributions.}:
\begin{gather*}
\zeta =1/\nu \ \rightarrow 0\\
K( -1/\zeta ) =\sqrt{\zeta }\left( -\frac{K(-\zeta )}{\pi }\textcolor[rgb]{0.82,0.01,0.11}{\ln}\textcolor[rgb]{0.82,0.01,0.11}{\zeta } +f_{\text{reg} |K}( \zeta )\right) ,\ \\
E( -1/\zeta ) =\frac{1}{\sqrt{\zeta }}\left( -\frac{(\zeta +1)K(-\zeta )-E(-\zeta )}{\pi }\textcolor[rgb]{0.82,0.01,0.11}{\ln}\textcolor[rgb]{0.82,0.01,0.11}{\zeta \ } +f_{\text{reg} |E}( \zeta )\right) .
\end{gather*}
For practical purposes it is also useful to indicate the local behavior of all regular expressions:
\begin{gather*}
K(-\zeta )\rightarrow \pi /2,\ E( -\zeta )\rightarrow \pi /2\\
f_{\text{reg} |K}( \zeta )\rightarrow \ln 4+\zeta \frac{1}{4} (1-\ln 4)+\dotsc \ ,\ f_{\text{reg} |E}( \zeta ) =1+\zeta \left(\frac{1}{4} +\ln 2\right) +\dotsc 
\end{gather*}

\subsection{Torus topology}
For simplicity, we first present a simple method for very accurate integration of the response in a torus-type topology.
Although such a setting is exotic for vesicles, it is fully admissible \cite{Seifert1991a}.
Moreover, this setting naturally arises when modeling a nozzle of a given shape.

In this case, it suffices to subtract the logarithmic singularity in a periodic form, after which all regular functions become periodic, and then integration via the Fourier transform is exponentially accurate.

\subsubsection{Extracting a convenient logarithm}

Since the logarithmic part has the form
\begin{equation}
f( \tau )\ln \zeta ( \tau )
\end{equation}
with smooth $\displaystyle f( \tau ) ,\ \zeta ( \tau )$ and a single second-order zero in $\displaystyle \zeta \left( \tau \rightarrow \tau _{o}\right)$, one needs to extract the singularity explicitly in a convenient way. We propose to use, as an integrand regularizer,
\begin{equation}
\aleph ( \tau ) \ =\frac{\sin^{2}( \pi ( \tau -\tau _{o}))}{2}
\end{equation}
Then the original function can be represented as
\begin{equation}
f( \tau )\ln \zeta ( \tau ) \ =\ f( \tau )\ln\frac{\zeta }{\aleph } + f( \tau )\textcolor[rgb]{0.82,0.01,0.11}{\ln}\textcolor[rgb]{0.82,0.01,0.11}{\aleph } ,
\end{equation}
where the singular part is made explicit, and the first term is already smooth.\footnote{Of course, any convenient $\displaystyle \aleph $ that keeps $\displaystyle \zeta /\aleph $ away from zero is suitable.}
\begin{equation*}
\zeta /\aleph \rightarrow \frac{1}{2\rho _{o}^{2}}\frac{V_{o}^{2}}{\pi ^{2}}
\end{equation*}
\subsubsection{Quadrature rule}

Since the integrand (after shifting) has the form
\begin{equation}
f\ =\ \ln\frac{\sin^{2}( \pi \tau )}{2} f_{\text{reg}}( \tau ) ,
\end{equation}
for the factor $\displaystyle f_{\text{reg}}( \tau )$ that is smooth on the circle, the best analytic continuation from discrete points is simply the Fourier transform on a uniform grid. 

Thus, the problem reduces to taking the discrete Fourier transform and using the basic integrals:
\begin{gather}
\int _{0}^{1} d\tau \ln\frac{\sin^{2}( \pi \tau )}{2} \ =\ -\ln 8\\
\int _{0}^{1} d\tau \ln\frac{\sin^{2}( \pi \tau )}{2}\cos( 2\pi k\tau ) \ =\ -\frac{1}{k}\\
\int _{0}^{1} d\tau \ln\frac{\sin^{2}( \pi \tau )}{2}\sin( 2\pi k\tau ) =0
\end{gather}

\begin{figure}[H]
	\centering
	\begin{subfigure}[b]{0.15\textwidth}
		\centering
		\includegraphics[width=\textwidth]{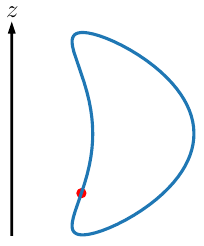}\vspace{6ex}
		\caption{}
		\label{fig:Toroid_shape_for_G_zz_demo}
	\end{subfigure}\hspace{3ex}
	\begin{subfigure}[b]{0.35\textwidth}
		\centering
		\includegraphics[width=\textwidth]{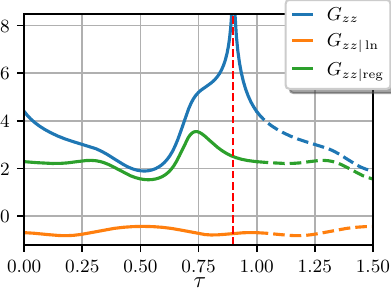}
		\caption{}
		\label{fig:Toroid_G_zz_demo}
	\end{subfigure}\hspace{3ex}
	\begin{subfigure}[b]{0.35\textwidth}
		\centering
		\includegraphics{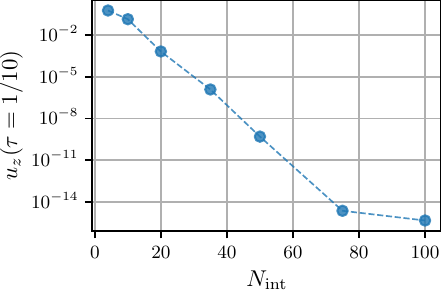}
		\caption{}
		\label{fig:error_u_z_toroid_tau_1_10}
	\end{subfigure}
	\caption{Demonstration of an exponentially accurate scheme for integrating the Green's function on a toroidal vesicle. For the vesicle shape (\subref{fig:Toroid_shape_for_G_zz_demo}), panel \subref{fig:Toroid_G_zz_demo} shows the Green's function $G_{zz}(\tau_o, \tau)$ for computing the velocity at the red point (\subref{fig:Toroid_shape_for_G_zz_demo}), together with the prefactor multiplying the logarithm and the regular remainder. Panel (\subref{fig:error_u_z_toroid_tau_1_10})} demonstrates exponential error decay as the number of quadrature points increases.
	\label{fig:toroid_demo}
\end{figure}

Figure \ref{fig:toroid_demo} shows components of the Green's function and the error in computing the $z$ component of the velocity at a selected point on the vesicle as a function of the number of uniform quadrature points (for simplicity, we compute the velocity induced by a constant normal force on the vesicle; the induced velocity vanishes due to cancellation by the pressure jump).

\subsection{Methods for sphere topology}
\begin{figure}
		\centering
		\includegraphics[width=0.35\textwidth]{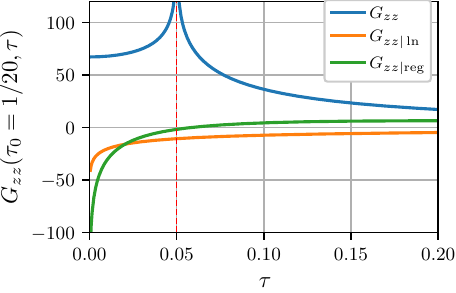}
	\caption{A Green's-function component on the surface (\ref{fig:Shape_for_curvature_demo}) $G_{zz}(\tau_o=1/20, tau)$ with the locally regular prefactor multiplying the logarithm and the regular remainder. One can see that these locally regular parts have a singularity on the axis.}    
			\label{fig:G_zz_with_reg_heart}
\end{figure}

Note that, for sphere topology, there is an axis—an additional singularity. 
As can be shown straightforwardly, in this case a globally smooth separation of the response into a logarithmic part and a regular part is impossible on the full interval: the regular part—the logarithmic prefactor—diverges on the axis (see Fig. \ref{fig:G_zz_with_reg_heart}), which makes the method described above inapplicable. We emphasize that this feature, strictly speaking, prevents one from using the Kapur--Rokhlin and Alpert methods on the entire interval, since they assume a smooth separation.

\begin{wrapfigure}[4]{l}{0.3\textwidth}
	\begin{center}
		\vspace{-6ex}
		\includegraphics[width=0.3\textwidth]{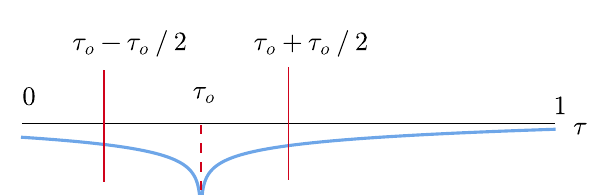}
	\end{center}
\end{wrapfigure}Therefore, one has to split the full integration interval into a part near the singularity and the remaining parts.
For example, one may simply split it in half with respect to the nearest boundary.
The most straightforward stable high-order integration on a subinterval using a uniform grid can be performed using a generalization of the Maclaurin--Euler formula to functions with a logarithmic factor\cite{Navot1962}.

\begin{equation*}
f( x) =g( x)\ln x
\end{equation*}
Then the integral
\begin{gather*}
\sum _{j=0}^{n-1} f\left(\frac{j+a}{n}\right) -n\int _{0}^{1} f( x) dx\ =\ \\
\sum _{k=1}^{k=2m-1}\frac{B_{k}( a)}{k!}\frac{1}{n^{k-1}}\left\{f^{( k-1)}( 1) +\ln n\ g^{( k-1)}( 0)\right\} +\frac{B_{2m}( a)}{( 2m) !}\left(\frac{1}{n}\right)^{2m-1}\ln n\ g^{( 2m-1)}( 0)
-\sum _{k=0}^{2m-2} \zeta '( -k,a)\left(\frac{1}{n}\right)^{k}\frac{1}{k!} g^{( k)}( 0)
\end{gather*}

\subsubsection{Smooth localization of the singularity}

\begin{wrapfigure}[8]{l}{0.3\textwidth}
	\begin{center}
		\vspace{-5ex}
		\includegraphics[width=0.3\textwidth]{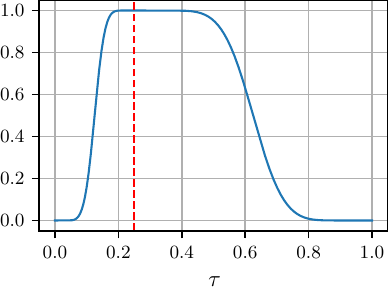}
	\end{center}
\end{wrapfigure} We can smoothly localize the problematic part, so that one only needs to integrate a smooth (but non-analytic) function outside the neighborhood, and a smooth function weighted by a logarithm inside, which can also be handled explicitly as in the torus-topology case.
\begin{gather*}
\eta ( x) =e^{-\frac{1}{x}}\mathbb{I}_{x >0}\\
s_{l}( \tau ) \ =\ \frac{\eta ( 1-\tau /\tau _{o})}{\eta ( 1-\tau /\tau _{o}) +\eta ( \tau /\tau _{o})}
\end{gather*}.

\subsubsection{Subtracting the singularity + $C_1$ local reconstruction}

An alternative approach, the simplest to implement, is to eliminate the logarithmic singularity by direct subtraction. 
\begin{equation*}
\int _{0}^{1} d\tau _{1} f( \tau ,\tau _{1}) ,\ f\sim C_{\ln}( \tau )\ln |\tau -\tau _{1} |\ +C_{0}( \tau ) +\ \mathcal{O}( \tau -\tau _{1})
\end{equation*}

The simplest method is to subtract the leading singularity, which can be integrated analytically:
\begin{gather*}
\int _{0}^{1} d\tau _{1} f( \tau ,\tau _{1}) \ =\ \int _{0}^{1} d\tau _{1}\tilde{f}( \tau ,\tau _{1}) +C_{\ln}( \tau ) I_{\ln}( \tau ) ,\\
\tilde{f}( \tau ,\tau _{1}) =f( \tau ,\tau _{1}) -C_{\ln}( \tau )\ln |\tau -\tau _{1} |,\ \tilde{f}( \tau ,\tau ) =C_{0}( \tau )\\
I_{\ln}( \tau ) =\int _{0}^{1} d\tau _{1}\ln |\tau -\tau _{1} |\ =\ (1-\tau )\ln (1-\tau )+\tau \log (\tau )-1.
\end{gather*}
Note that $\displaystyle \tilde{f}$ is continuous; however, near $\displaystyle \tau _{1} \approx \tau $ there still remains a singularity of the form $\displaystyle ( \tau -\tau _{1})\ln |\tau -\tau _{1} |$. The simplest next step is to integrate $\displaystyle \tilde{f}$ by standard methods; as an example, consider the trapezoidal rule, where a piecewise-linear reconstruction is performed from the given nodes, followed by integration.

Note that the naive error estimate for such a procedure with step size $\displaystyle \epsilon $
\begin{equation*}
Res\sim \tilde{f} ''( \xi ) \epsilon ^{2} ,\ \xi \in [ 0,1]
\end{equation*}
is a strong overestimate if one assumes that the error accumulates near the singularity where \ $\displaystyle \tilde{f} ''\sim \frac{1}{\tau -\tau _{1}} \sim \frac{1}{\epsilon }$,
so that $\displaystyle Res\sim \epsilon $. The point is that the integral of the leading singularity over its neighborhood is zero by symmetry, which the trapezoidal rule also reproduces when using uniformly spaced quadrature nodes; thus one must consider the next non-analytic term in the expansion, $\displaystyle ( \tau -\tau _{1})^{2}\ln |\tau -\tau _{1} |$. Integrating it by the trapezoidal rule yields an error $\displaystyle \sim \epsilon ^{2}$ (an estimate based on a characteristic derivative gives an overestimate, since the result depends on its integral, and the logarithmic singularity is integrable). The resulting estimate makes it possible to improve the asymptotics at least to $\displaystyle \epsilon ^{3}\ln \epsilon $ by switching to Simpson's rule (which corresponds to one step of the Runge rule).

A more efficient approach is the following: based on the error estimate, one can propose a smoother ($\displaystyle C_{1}$) reconstruction of the integrand. We use the previously derived finite elements, which lead to the following quadrature rule:
\begin{equation}
\int _{0}^{1}\tilde{f}\rightarrow \frac{9}{24} f_{0} +\frac{28}{24} f_{1} +\frac{23}{24} f_{2} +f_{3} +\dotsc +f_{-4} +\frac{23}{24} f_{-3} +\frac{28}{24} f_{-2} +\frac{9}{24} f_{-1} =\sum _{j=0}^{j=N_{int}} w_{j}\tilde{f}_{j}
\end{equation}
Let us present the full form of the method:
\begin{gather*}
\int _{0}^{1} d\tau _{1} f( \tau ,\tau _{1}) \ =w_{j}\tilde{f}_{j}( \tau ) \ +C_{\ln}( \tau ) I_{\ln}( \tau ) =\\
w_{j} \delta _{\tau _{j} \neq \tau } f_{j}( \tau ) -C_{\ln}( \tau ) w_{j} \delta _{\tau _{j} \neq \tau }\ln |\tau -\tau _{j} |+w_{j:\tau _{j} =\tau } C_{0}( \tau ) +C_{\ln}( \tau ) I_{\ln}( \tau ) =\\
w_{j} \delta _{\tau _{j} \neq \tau } f_{j}( \tau ) +w_{j:\tau _{j} =\tau } C_{0}( \tau ) +C_{\ln}( \tau )[ I_{\ln}( \tau ) -I_{\ln |\mathrm{quad}}( \tau )] ,\\
I_{\ln |\mathrm{quad}} \ =\ w_{j} \delta _{\tau _{j} \neq \tau }\ln |\tau -\tau _{j} |,\ I_{\ln |\delta }( \tau ) =\ I_{\ln}( \tau ) -I_{\ln |\mathrm{quad}}( \tau )\\
\boxed{\int _{0}^{1} d\tau _{1} f( \tau ,\tau _{1}) \ \ =\ w_{j} \delta _{\tau _{j} \neq \tau } f_{j}( \tau ) +w_{j:\tau _{j} =\tau } C_{0}[ f]( \tau ) +C_{\ln}[ f]( \tau ) I_{\ln |\delta }( \tau )}
\end{gather*}

\subsubsection{Comparison of integration methods}

\begin{figure}[H]
	\centering
	\begin{subfigure}[b]{0.3\textwidth}
		\centering
		\includegraphics[width=\textwidth]{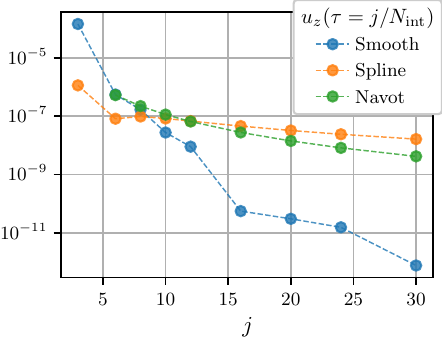}
		\caption{}
		\label{fig:error_u_z_heart_j_ov_2400}
	\end{subfigure}\
	\begin{subfigure}[b]{0.3\textwidth}
		\centering
		\includegraphics[width=\textwidth]{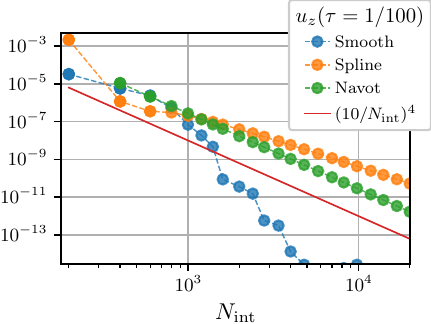}
		\caption{}
		\label{fig:error_u_z_heart_tau_1_100}
	\end{subfigure}
	\begin{subfigure}[b]{0.3\textwidth}
		\centering
		\includegraphics[width=\textwidth]{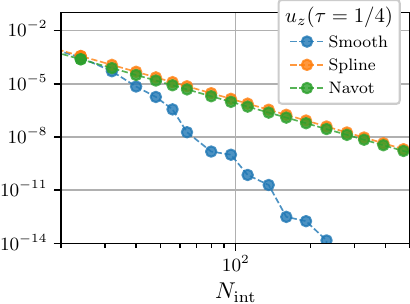}
		\caption{}
		\label{fig:error_u_z_heart_tau_1_4}
	\end{subfigure}
	\caption{
		(\subref{fig:error_u_z_heart_j_ov_2400}) Dependence of the error in computing $u_z$ at node number $j$, with the total number of quadrature points $N_\mathrm{int}=2400$.
		(\subref{fig:error_u_z_heart_tau_1_100}) Shows the accuracy of each method for a node near the symmetry axis $\tau_o=1/100$. We see that the advantage of the Smooth method is substantial for $N_\mathrm{int}\tau_o\approx 20$.
		(\subref{fig:error_u_z_heart_tau_1_100}) Demonstrates the same, but for a point far from the axis.
}
	\label{fig:comparing_mathods_int}
\end{figure}

In Fig. \ref{fig:comparing_mathods_int} we compare the errors of different methods (Navot, fourth order).
In Fig. \ref{fig:error_u_z_heart_j_ov_2400} we see that the accuracy of each method decreases as the evaluation point approaches the symmetry axis. Figures \ref{fig:error_u_z_heart_tau_1_100} and \ref{fig:error_u_z_heart_tau_1_100} confirm the advantage of smooth localization over the other methods. We note, however, that for a sufficiently large number of points ($N_\mathrm{int} \tau_o\sim 20$) all methods already yield a good result with relative error $10^{-7}$. The spline-based method is about three times faster, since it does not require splitting the integration intervals and recomputing special functions.
Therefore, for modest accuracy requirements we recommend using it, and if that is insufficient, the Smooth method.

We additionally note that the Smooth and Spline methods are also applicable to computing the surface divergence, where the leading singularity has the form $1/x$, which, together with their simplicity, is a major advantage.
\section{Illustrative experiments}
Here, after the publication of certain physical results\cite{Shishkin2025}, examples of full dynamical simulations will be added.

\section{Conclusion}
The main results are the improvements of \cite{Veerapaneni2009} listed above.
\begin{itemize}
	\item Construction of a simple reparameterization procedure based on a local characteristic length scale, which makes it possible to use a smaller number of harmonics for the same accuracy
	\item Construction of a gauge dynamics: a method for selecting the tangential velocity from the normal one so that the parameterization remains optimal
	\item Development of a scheme for preventing parametric growth of the numerical error in the computation of physical fields such as the Helfrich force near the symmetry axis
	\item Construction of a spectrally accurate quadrature scheme for evaluating integrals with singularities (based on an analytically derived singular decomposition of the axial Green's function) 
\end{itemize}

 \section*{Acknowledgment}

M.A.Sh. gratefully acknowledges the Foundation for the Advancement of Theoretical Physics and Mathematics ``BASIS'' for support.

\nocite{*}
\printbibliography

@article{Alpert1999,
  title = {Hybrid {{Gauss-Trapezoidal Quadrature Rules}}},
  author = {Alpert, Bradley K.},
  date = {1999-01},
  journaltitle = {SIAM Journal on Scientific Computing},
  shortjournal = {SIAM J. Sci. Comput.},
  volume = {20},
  number = {5},
  pages = {1551--1584},
  issn = {1064-8275, 1095-7197},
  doi = {10.1137/S1064827597325141},
  url = {http://epubs.siam.org/doi/10.1137/S1064827597325141},
  urldate = {2026-04-14},
  langid = {english}
}

@online{Ambrose2021,
  title = {Convergence of the Boundary Integral Method for Interfacial {{Stokes}} Flow},
  author = {Ambrose, David M. and Siegel, Michael and Zhang, Keyang},
  date = {2021-05-14},
  eprint = {2105.07056},
  eprinttype = {arXiv},
  eprintclass = {math},
  doi = {10.48550/arXiv.2105.07056},
  url = {http://arxiv.org/abs/2105.07056},
  urldate = {2026-04-14},
  pubstate = {prepublished}
}

@online{Hao2012,
  title = {High-Order Accurate {{Nystrom}} Discretization of Integral Equations with Weakly Singular Kernels on Smooth Curves in the Plane},
  author = {Hao, S. and Barnett, A. H. and Martinsson, P. G. and Young, P.},
  date = {2012-11-21},
  eprint = {1112.6262},
  eprinttype = {arXiv},
  eprintclass = {math},
  doi = {10.48550/arXiv.1112.6262},
  url = {http://arxiv.org/abs/1112.6262},
  urldate = {2025-12-10},
  langid = {english},
  pubstate = {prepublished}
}

@article{Heintz2015,
  title = {A Numerical Method for Simulation Dynamics of Incompressible Lipid Membranes in Viscous Fluid},
  author = {Heintz, A.},
  date = {2015-12},
  journaltitle = {Journal of Computational and Applied Mathematics},
  shortjournal = {Journal of Computational and Applied Mathematics},
  volume = {289},
  pages = {87--100},
  issn = {03770427},
  doi = {10.1016/j.cam.2015.03.011},
  url = {https://linkinghub.elsevier.com/retrieve/pii/S0377042715001570},
  urldate = {2026-04-14},
  langid = {english}
}

@article{Helfrich1973,
  title = {Elastic {{Properties}} of {{Lipid Bilayers}}: {{Theory}} and {{Possible Experiments}}},
  author = {Helfrich, W.},
  date = {1973-12-01},
  journaltitle = {Zeitschrift für Naturforschung C},
  volume = {28},
  number = {11--12},
  eprint = {4273690},
  eprinttype = {pubmed},
  pages = {693--703},
  issn = {1865-7125},
  doi = {10.1515/znc-1973-11-1209},
  url = {https://www.degruyter.com/document/doi/10.1515/znc-1973-11-1209/html}
}

@article{Hu2014,
  title = {An Immersed Boundary Method for Simulating the Dynamics of Three-Dimensional Axisymmetric Vesicles in {{Navier}}–{{Stokes}} Flows},
  author = {Hu, Wei-Fan and Kim, Yongsam and Lai, Ming-Chih},
  date = {2014-01},
  journaltitle = {Journal of Computational Physics},
  shortjournal = {Journal of Computational Physics},
  volume = {257},
  pages = {670--686},
  issn = {00219991},
  doi = {10.1016/j.jcp.2013.10.018},
  url = {https://linkinghub.elsevier.com/retrieve/pii/S0021999113006943},
  urldate = {2026-04-14},
  langid = {english}
}

@article{Kapur1997,
  title = {High-{{Order Corrected Trapezoidal Quadrature Rules}} for {{Singular Functions}}},
  author = {Kapur, Sharad and Rokhlin, Vladimir},
  date = {1997-08},
  journaltitle = {SIAM Journal on Numerical Analysis},
  shortjournal = {SIAM J. Numer. Anal.},
  volume = {34},
  number = {4},
  pages = {1331--1356},
  issn = {0036-1429, 1095-7170},
  doi = {10.1137/S0036142995287847},
  url = {http://epubs.siam.org/doi/10.1137/S0036142995287847},
  urldate = {2025-12-09},
  langid = {english}
}

@article{Keaveny2011,
  title = {Applying a Second-Kind Boundary Integral Equation for Surface Tractions in {{Stokes}} Flow},
  author = {Keaveny, Eric E. and Shelley, Michael J.},
  date = {2011-03},
  journaltitle = {Journal of Computational Physics},
  shortjournal = {Journal of Computational Physics},
  volume = {230},
  number = {5},
  pages = {2141--2159},
  issn = {00219991},
  doi = {10.1016/j.jcp.2010.12.010},
  url = {https://linkinghub.elsevier.com/retrieve/pii/S0021999110006741},
  urldate = {2026-04-14},
  langid = {english}
}

@incollection{Komzsik2020,
  title = {Differential Geometry},
  booktitle = {Applied {{Calculus}} of {{Variations}} for {{Engineers}}},
  author = {Komzsik, Louis},
  date = {2008-10-27},
  pages = {101--114},
  publisher = {CRC Press},
  doi = {10.1201/9781420086652-15},
  url = {https://www.taylorfrancis.com/books/9781420086652/chapters/10.1201/9781420086652-15},
  isbn = {978-3-031-39837-7}
}

@article{Lai2022,
  title = {A Stable and Accurate Immersed Boundary Method for Simulating Vesicle Dynamics via Spherical Harmonics},
  author = {Lai, Ming-Chih and Seol, Yunchang},
  date = {2022-01},
  journaltitle = {Journal of Computational Physics},
  shortjournal = {Journal of Computational Physics},
  volume = {449},
  pages = {110785},
  issn = {00219991},
  doi = {10.1016/j.jcp.2021.110785},
  url = {https://linkinghub.elsevier.com/retrieve/pii/S002199912100680X},
  urldate = {2026-04-14},
  langid = {english}
}

@article{Li2011,
  title = {New {{Finite Difference Methods Based}} on {{IIM}} for {{Inextensible Interfaces}} in {{Incompressible Flows}}},
  author = {Li, Zhilin and Lai, Ming-Chih},
  date = {2011-05},
  journaltitle = {East Asian Journal on Applied Mathematics},
  shortjournal = {E Asian J Appl Math},
  volume = {1},
  number = {2},
  pages = {155--171},
  issn = {2079-7362, 2079-7370},
  doi = {10.4208/eajam.030510.250910a},
  url = {https://www.cambridge.org/core/product/identifier/S2079736200000109/type/journal_article},
  urldate = {2026-04-14},
  langid = {english}
}

@article{Navot1962,
  title = {A {{Further Extension}} of the {{Euler}}‐{{Maclaurin Summation Formula}}},
  author = {Navot, Israel},
  date = {1962-04},
  journaltitle = {Journal of Mathematics and Physics},
  shortjournal = {Journal of Mathematics and Physics},
  volume = {41},
  number = {1--4},
  pages = {155--163},
  issn = {0097-1421},
  doi = {10.1002/sapm1962411155},
  url = {https://onlinelibrary.wiley.com/doi/10.1002/sapm1962411155},
  urldate = {2026-02-09},
  langid = {english}
}

@book{Pozrikidis1992,
  title = {Boundary {{Integral}} and {{Singularity Methods}} for {{Linearized Viscous Flow}}},
  author = {Pozrikidis, C.},
  date = {1992},
  doi = {10.1017/cbo9780511624124},
  isbn = {978-0-521-40502-7}
}

@article{Pozrikidis2001,
  title = {Interfacial {{Dynamics}} for {{Stokes Flow}}},
  author = {Pozrikidis, C.},
  date = {2001-05},
  journaltitle = {Journal of Computational Physics},
  shortjournal = {Journal of Computational Physics},
  volume = {169},
  number = {2},
  pages = {250--301},
  issn = {00219991},
  doi = {10.1006/jcph.2000.6582},
  url = {https://linkinghub.elsevier.com/retrieve/pii/S0021999100965823},
  urldate = {2025-12-08},
  langid = {english}
}

@book{Press2007,
  title = {Numerical Recipes: The Art of Scientific Computing},
  shorttitle = {Numerical Recipes},
  editor = {Press, William H.},
  date = {2007},
  edition = {3rd ed},
  publisher = {Cambridge University Press},
  location = {Cambridge, UK ; New York},
  isbn = {978-0-521-88068-8 978-0-521-88407-5 978-0-521-70685-8},
  pagetotal = {1235}
}

@article{Salac2011,
  title = {A Level Set Projection Model of Lipid Vesicles in General Flows},
  author = {Salac, D. and Miksis, M.},
  date = {2011-09},
  journaltitle = {Journal of Computational Physics},
  shortjournal = {Journal of Computational Physics},
  volume = {230},
  number = {22},
  pages = {8192--8215},
  issn = {00219991},
  doi = {10.1016/j.jcp.2011.07.019},
  url = {https://linkinghub.elsevier.com/retrieve/pii/S0021999111004384},
  urldate = {2026-04-14},
  langid = {english}
}

@article{Seifert1991a,
  title = {Vesicles of Toroidal Topology},
  author = {Seifert, Udo},
  date = {1991-05-06},
  journaltitle = {Physical Review Letters},
  shortjournal = {Phys. Rev. Lett.},
  volume = {66},
  number = {18},
  pages = {2404--2407},
  issn = {0031-9007},
  doi = {10.1103/PhysRevLett.66.2404},
  url = {https://link.aps.org/doi/10.1103/PhysRevLett.66.2404},
  urldate = {2025-12-17},
  langid = {english}
}

@online{Shishkin2025,
  title = {The Metastability of Lipid Vesicle Shapes in Uniaxial Extensional Flow},
  author = {Shishkin, M. A. and Pikina, E. S.},
  date = {2025},
  doi = {10.48550/ARXIV.2511.20840},
  url = {https://arxiv.org/abs/2511.20840},
  urldate = {2026-04-13},
  pubstate = {prepublished},
  version = {1}
}

@article{Singh2012,
  title = {A {{New Boundary Integral Formulation}} for {{Stream Function}} and {{Vorticity}} in {{Axisymmetric Stokes Flow}}},
  author = {Singh, Jitendra},
  date = {2012-01},
  journaltitle = {SIAM Journal on Applied Mathematics},
  shortjournal = {SIAM J. Appl. Math.},
  volume = {72},
  number = {5},
  pages = {1575--1591},
  issn = {0036-1399, 1095-712X},
  doi = {10.1137/110855922},
  url = {http://epubs.siam.org/doi/10.1137/110855922},
  urldate = {2026-04-14},
  langid = {english}
}

@article{Spann2014a,
  title = {Loop Subdivision Surface Boundary Integral Method Simulations of Vesicles at Low Reduced Volume Ratio in Shear and Extensional Flow},
  author = {Spann, Andrew P. and Zhao, Hong and Shaqfeh, Eric S. G.},
  date = {2014-03-01},
  journaltitle = {Physics of Fluids},
  volume = {26},
  number = {3},
  pages = {031902},
  issn = {1070-6631, 1089-7666},
  doi = {10.1063/1.4869307},
  url = {https://pubs.aip.org/pof/article/26/3/031902/258709/Loop-subdivision-surface-boundary-integral-method},
  urldate = {2026-04-14},
  langid = {english}
}

@article{Tan2012,
  title = {An {{Immersed Interface Method}} for the {{Simulation}} of {{Inextensible Interfaces}} in {{Viscous Fluids}}},
  author = {Tan, Zhijun and Le, D. V. and Lim, K. M. and Khoo, B. C.},
  date = {2012-03},
  journaltitle = {Communications in Computational Physics},
  shortjournal = {Commun. comput. phys.},
  volume = {11},
  number = {3},
  pages = {925--950},
  issn = {1815-2406, 1991-7120},
  doi = {10.4208/cicp.200110.040511a},
  url = {https://www.cambridge.org/core/product/identifier/S1815240600002528/type/journal_article},
  urldate = {2026-04-14},
  langid = {english}
}

@article{Veerapaneni2009,
  title = {A Numerical Method for Simulating the Dynamics of {{3D}} Axisymmetric Vesicles Suspended in Viscous Flows},
  author = {Veerapaneni, Shravan K. and Gueyffier, Denis and Biros, George and Zorin, Denis},
  date = {2009},
  journaltitle = {Journal of Computational Physics},
  volume = {228},
  number = {19},
  pages = {7233--7249},
  publisher = {Elsevier Inc.},
  issn = {10902716},
  doi = {10.1016/j.jcp.2009.06.020},
  url = {http://dx.doi.org/10.1016/j.jcp.2009.06.020},
  isbn = {2129983510},
  langid = {american}
}

@article{Veerapaneni2011,
  title = {A Fast Algorithm for Simulating Vesicle Flows in Three Dimensions},
  author = {Veerapaneni, Shravan K. and Rahimian, Abtin and Biros, George and Zorin, Denis},
  date = {2011-06},
  journaltitle = {Journal of Computational Physics},
  shortjournal = {Journal of Computational Physics},
  volume = {230},
  number = {14},
  pages = {5610--5634},
  issn = {00219991},
  doi = {10.1016/j.jcp.2011.03.045},
  url = {https://linkinghub.elsevier.com/retrieve/pii/S0021999111002063},
  urldate = {2026-04-14},
  langid = {english}
}

@online{Wu2021,
  title = {Zeta {{Correction}}: {{A New Approach}} to {{Constructing Corrected Trapezoidal Quadrature Rules}} for {{Singular Integral Operators}}},
  shorttitle = {Zeta {{Correction}}},
  author = {Wu, Bowei and Martinsson, Per-Gunnar},
  date = {2021-04-07},
  eprint = {2007.13898},
  eprinttype = {arXiv},
  eprintclass = {math},
  doi = {10.48550/arXiv.2007.13898},
  url = {http://arxiv.org/abs/2007.13898},
  urldate = {2026-02-09},
  langid = {english},
  pubstate = {prepublished}
}

\clearpage
\appendix

\newcommand{\appsection}[1]{\let\oldthesection\thesection
	\renewcommand{\thesection}{Appendix \oldthesection}
	\section{#1}\let\thesection\oldthesection}

\phantomsection
\addcontentsline{toc}{section}{{Appendices}}

\appsection{Analytic continuation on the circle}
\label{pril_1}
Let us explicitly describe the procedure for analytic continuation of a periodic function from a set of points.
Due to the uniformity of the nodes and periodic boundary conditions, the analytic continuation of a function given at $\displaystyle n_{int}$ points on a circle of length 1 $\displaystyle \left( \tau _{i} =\frac{1}{n_{int}}\right)$ takes the form of a convolution with the response function:
\begin{gather*}
f_{cont}( \tau ) =\sum _{i} f_{i} R_{n_{int}}( \tau -\tau _{i}) ,\\
R_{n_{int}}( x) =\frac{\sin( \pi xn_{int})}{\pi xn_{int}}( \pi x\cdot \cot( \pi x)) =\text{sinc}( \pi xn_{int})\frac{\cos( \pi x)}{\text{sinc}( \pi x)} .
\end{gather*}
The response function is accordingly periodic and vanishes at all nodes not congruent to zero.
\begin{figure}
    \centering
    \includegraphics[width=1.\linewidth]{032_pic+Kotelnicov.png}
    \caption{Comparison of the response functions on the line $\text{sinc}(\pi x n_{int})$ (green) and on the circle (blue) for $n_{int}=30$.}
    \label{fig:enter-label}
\end{figure}
This formula can be obtained by direct resummation of the Kotelnikov formula for a periodic function. The figure below compares the response function on the circle with the response on the line. Note that, in both cases, the kernel decays algebraically with the node index, which is related to the infinite smoothness of the reconstruction.

The obtained relations are used to prescribe the vesicle state manually (see the interactive plot in \href{https://www.desmos.com/calculator/aiarx0n6ri}{Desmos}): by fixing a small number of nodes in the plane, we obtain a smooth curve $\displaystyle \mathbf{R}( \tau )$.\footnote{The task of drawing a smooth curve through a prescribed set of points is usually addressed in vector graphics using piecewise-smooth splines, which is inapplicable for prescribing the initial state of a smooth vesicle surface. The procedure described above makes it possible to construct an infinitely smooth reconstruction of closed curves, which may also be more convenient in applications.}
\appsection{Re-gauging procedure}
\label{app:regauge}
Suppose we are given a state $\displaystyle \mathbf{R}( \tau )$, and we want to obtain its representation in another gauge $\displaystyle \mathbf{R} '( \tau ')$, which is uniquely specified by the speed $\displaystyle V'( \tau ')$.
Using the transformation rule for the traversal speed under reparameterization, we obtain a simple defining equation if $\displaystyle V_d$ is a scalar field, i.e., $\displaystyle V_d'=V_d$:

\begin{equation*}
V'( \tau ') =CV_d'( \tau ')=V\frac{d\tau }{d\tau '} \Longrightarrow \boxed{\frac{d\tau '}{d\tau } =\frac{V( \tau )}{CV_d'( \tau ')} =\frac{1}{C}\frac{V( \tau )}{V_d( \tau )}}
\end{equation*}

It is convenient to solve this equation in Fourier space, using the expansion of $V(\tau)$:
\begin{gather*}
\tau '(\tau)=\tau +\sum _{k} c_{k} \varphi _{\rho k}\implies\\
\frac{d\tau '}{d\tau } \ =\ 1+\sum _{k=1} c_{k}( \pi k) \varphi _{zk} =\frac{1}{C}\left(\sum _{k=0} \Vert V/V_d\Vert_k \varphi _{zk}\right) \Longrightarrow \\
\begin{cases}
C=\Vert V/V_d\Vert_{0} & \\
c_{k} =\frac{1}{\pi k}\frac{1}{C}\Vert V/V_d\Vert_k , & k\geqslant 1
\end{cases}
\end{gather*}
Thus, we determine $\tau'(\tau)$, and the new representation is obtained by substituting the inverse transformation:
\begin{equation}
\mathbf{R}'(\tau')=\mathbf{R}(\tau(\tau')).
\end{equation}

Note that, taken literally, such a transformation requires solving a nonlinear problem; therefore, we indicate a more convenient approach for practical implementation:
in any case, we will have to perform not only a gauge change, but also the projection onto a finite-dimensional basis $\mathbf{R}'_N[\{p_k, z_k\}]$:
\begin{equation}
\Vert \mathbf{R}'_N(\tau')-\mathbf{R}'(\tau')\Vert_{L_2(d\tau')}\to\min .
\end{equation}
The corresponding extremal problem for the new expansion coefficients can be rewritten in terms of integrals in the original gauge:
\begin{gather}
\int  \left(\sum _k p_k'\varphi_{\rho k}(\tau')-P'(\tau')\right)^2 d\tau'= \\
\int  \left(\sum _k p_k'\varphi_{\rho k}(\tau'(\tau))-P(\tau)\right)^2 \underbrace{{d\tau'}/{d\tau}}_{\propto V/V_d} \; d\tau,
\end{gather}
which, taking into account the uniformity of the points $\tau_j$, yields an LLS problem for the vectors $\{p_k\}$(see. \ref{quad_approx}):
\begin{equation}
\sum_k p_k'\; \sqrt{(V/V_d)(\tau_j)} \varphi_{\rho k}(\tau'(\tau_j))\approx P(\tau_j)\sqrt{(V/V_d)(\tau_j)}.
\end{equation}

\appsection{Expressions for physical quantities in an arbitrary parameterization}
Let us express the various physical quantities we require in a general parameterization $\displaystyle \mathbf{R}( \tau )$.

The first fundamental form of the surface is expressed via the metric in cylindrical coordinates $\displaystyle ( \rho ,z,\varphi )$:
\begin{equation*}
ds^{2} \ =\ dz^{2} +d\rho ^{2} +\rho ^{2} d\varphi ^{2} \ =\ \underbrace{\left( Z_{\tau }^{2} +P_{\tau }^{2}\right)}_{v^{2}} d\tau ^{2} +P^{2} d\varphi ^{2} ,
\end{equation*}
where $v=\frac{dl}{d\tau}$ is the “traversal speed” along the curve.
Hence, the metric tensor in the $\displaystyle ( \tau ,\varphi )$ components is
\begin{equation*}
g_{ij} \ =\ \begin{pmatrix}
v^{2} \  & 0\\
0 & P^{2}
\end{pmatrix} \ \ g^{ij} =\begin{pmatrix}
\frac{1}{v^{2}} & 0\\
0 & \frac{1}{P^{2}}
\end{pmatrix} .
\end{equation*}
Accordingly, the area element is
\begin{equation*}
dS\ =\ vP\ d\tau d\varphi .
\end{equation*}
In particular, the surface area and vesicle volume are given by
\begin{equation*}
S\ =\ 2\pi \int _{0}^{1} vP\ d\tau ,\ V\ =\ \pi \int _{0}^{1} P^{2} Z_{\tau } d\tau .
\end{equation*}
The surface gradient operator is
\begin{equation*}
\nabla ^{\perp } \ =\ \mathbf{\tau}\frac{1}{v} \partial _{\tau } \ +\ \mathbf{e}_{\varphi }\frac{1}{P} \partial _{\varphi } ,
\end{equation*}
where the tangent vector is $\displaystyle \mathbf{\tau } \ =\frac{1}{v} \ \begin{pmatrix}
P_{\tau } & Z_{\tau }
\end{pmatrix}$ in the $\displaystyle ( \rho ,z)$ components. Let us also state the definition of the normal vector:
\begin{equation*}
\ \mathbf{l} \ =\ \frac{1}{v}\begin{pmatrix}
-Z_{\tau } & P_{\tau }
\end{pmatrix} .
\end{equation*}
To obtain the curvature radii, one has to compute the second (extrinsic) fundamental form\cite{Komzsik2020}:
\begin{gather*}
d\mathbf{r} \ =\ ( P_{\tau }\mathbf{e}_{\rho } \ +\ Z_{\tau }\mathbf{e}_{z}) d\tau \ +\ \mathbf{e}_{\varphi } Pd\varphi \\
d^{2}\mathbf{r} \ =\ ( P_{\tau \tau }\mathbf{e}_{\rho } +Z_{\tau \tau }\mathbf{e}_{z}) d\tau ^{2} \ +\ 2P_{\tau }\mathbf{e}_{\varphi } d\varphi d\tau \ \ -\mathbf{e}_{\rho } P\ d\varphi ^{2}\\
Q_{\tau \tau } \ =\ (\mathbf{l} ,\ ( P_{\tau \tau }\mathbf{e}_{\rho } +Z_{\tau \tau }\mathbf{e}_{z})) \ =\ -\frac{Z_{\tau }}{v} P_{\tau \tau } \ +\ \frac{P_{\tau }}{v} Z_{\tau \tau }\\
Q_{\varphi \varphi } \ =\ (\mathbf{l} ,\ -\mathbf{e}_{\rho } P) \ =\ \frac{Z_{\tau }}{v} ,\ Q_{\tau \varphi } =0\Longrightarrow Q=\begin{pmatrix}
\frac{P_{\tau } Z_{\tau \tau } -Z_{\tau } P_{\tau \tau }}{v} & \\
 & \frac{PZ_{\tau }}{v}
\end{pmatrix}\\
gQ\ =\ \begin{pmatrix}
\frac{P_{\tau } Z_{\tau \tau } -Z_{\tau } P_{\tau \tau }}{v^{3}} & \\
 & \frac{Z_{\tau }}{Pv}
\end{pmatrix} ,
\end{gather*}
whence we obtain the mean and Gaussian curvatures:
\begin{gather*}
-H\ =\ \text{Sp}( gQ) \ =\ \frac{P_{\tau } Z_{\tau \tau } -Z_{\tau } P_{\tau \tau }}{v^{3}} \ +\ \frac{Z_{\tau }}{Pv} ,\ K\ =\ \det( gQ) =\frac{P_{\tau } Z_{\tau \tau } -Z_{\tau } P_{\tau \tau }}{v^{3}}\frac{Z_{\tau }}{Pv}\\
H_{1} =\frac{Z_{\tau } P_{\tau \tau } -P_{\tau } Z_{\tau \tau }}{v^{3}} ,\ H_{2} =-\frac{Z_{\tau }}{Pv} .
\end{gather*}
In computations, the following convenient relations are often used:
\begin{gather*}
\mathbf{\tau } \ =\tau _{\rho }\mathbf{e}_{\rho } +\tau _{z}\mathbf{e}_{z} ;\ \\
\partial _{\varphi }\mathbf{\tau } \ =\tau _{\rho } \partial _{\varphi }\mathbf{e}_{\rho } =\tau _{\rho }\mathbf{e}_{\varphi }\\
\frac{1}{v} \partial _{\tau }\mathbf{\tau } \ =\frac{1}{v} \partial _{\tau }( \tau _{\rho }\mathbf{e}_{\rho } +\tau _{z}\mathbf{e}_{z}) =\frac{1}{v}(\mathbf{e}_{\rho } \partial _{\tau } \tau _{\rho } +\mathbf{e}_{z} \partial _{\tau } \tau _{z}) =-H_{1}\mathbf{l}\\
\mathnormal{\frac{1}{v} \partial _{\tau }}\mathbf{\ l} =H_{1}\mathbf{\tau }\\
\mathnormal{\partial _{\varphi }}\mathbf{l} \ =\ l_{\rho }\mathbf{e_{\varphi }}
\end{gather*}
Let us derive a coordinate expression for the Laplace--Beltrami operator:
\begin{equation*}
\left( \nabla ^{\perp } ,\nabla ^{\perp }\right) \ =\ \left(\mathbf{\tau }\frac{1}{v} \partial _{\tau } \ +\ \mathbf{e}_{\varphi }\frac{1}{P} \partial _{\varphi } ,\mathbf{\tau }\frac{1}{v} \partial _{\tau } \ +\ \mathbf{e}_{\varphi }\frac{1}{P} \partial _{\varphi }\right) =\frac{1}{vP}\left( \partial _{\tau }\frac{P}{v} \partial _{\tau }\right) +\frac{1}{P^{2}} \partial _{\varphi }^{2} =\Delta _{\tau }^{\perp } +\Delta _{\varphi }^{\perp}
\end{equation*}
A convenient form of the Laplacian operator applied to physical fields is
\begin{equation}
\frac{1}{vP}\left( \partial _{\tau }\frac{P}{v} \partial _{\tau }\right)H_2 = \frac{1}{v}\partial_\tau \frac{1}{v}\partial_\tau H_2+\frac{P'}{v^2}\frac{\partial_{\tau} H_2}{P}.
\end{equation}

These relations make it possible to obtain an explicit expression for the Helfrich force in an arbitrary parameterization in terms of the functions $P(\tau), Z(\tau)$ and their derivatives up to fourth order.

\appsection{Green axial}\label{app:green_axial}
Naturally, the Green's function can be expressed in terms of the planar distance $\displaystyle \delta r$ and the geometric parameter $\displaystyle \nu $.
\begin{equation*}
\delta r^{2} =z^{2} +( \rho -\rho _{1})^{2} ,\ \nu =\ \frac{4\rho \rho _{1}}{\delta r^{2}}
\end{equation*}
\begin{gather*}
G_{zz} =\frac{4}{\delta r}\left( K(-\nu )+\frac{z^{2}}{z^{2} +(\rho +\rho _{1} )^{2}} E(-\nu )\right)\\
G_{z\rho } =\frac{2}{\delta r}\frac{z}{\rho _{1}}\left( -K(-\nu )+\frac{z^{2} +\rho ^{2} -\rho _{1}^{2}}{z^{2} +(\rho +\rho _{1} )^{2}} E(-\nu )\right)\\
G_{\rho z} =-\frac{2}{\delta r}\frac{z}{\rho }\left( -K(-\nu )+\frac{z^{2} +\rho _{1}^{2} -\rho ^{2}}{z^{2} +(\rho +\rho _{1} )^{2}} E(-\nu )\right)\\
G_{\rho \rho } =\frac{2}{\delta r}\frac{1}{\rho \rho _{1}}\left( [\rho ^{2} +\rho _{1}^{2} +2z^{2} ]K(-\nu )-\frac{(\rho ^{2} -\rho _{1}^{2} )^{2} +3(\rho ^{2} +\rho _{1}^{2} )z^{2} +2z^{4}}{(\rho +\rho _{1} )^{2} +z^{2}} E(-\nu )\right) ,
\end{gather*}
where $\displaystyle K,\ E$ \ are elliptic functions:
\begin{gather*}
K(-\nu )=\int _{0}^{\pi /2} d\varphi \frac{1}{\sqrt{1+\nu \sin^{2} \varphi }} ,\ E(-\nu )=\int _{0}^{\pi /2} d\varphi \sqrt{1+\nu \sin^{2} \varphi } .
\end{gather*}

\begin{quote}
	{\color{red}Attention! Wolfram uses an unusual definition of the function $K$; in fact, it is the \href{https://en.wikipedia.org/wiki/Quarter_period}{Quarter period}, related to the complete elliptic integral of the first kind in the form $\text{Elliptic integral}(k) = K(k^2)$}
\end{quote}

Note that, using elliptic identities 
$$
K(-\nu) = \frac{K(\frac{\nu}{1+\nu})}{\sqrt{1+\nu}}, \; E(-\nu) = \sqrt{\nu+1}E(\frac{\nu}{1+\nu}),
$$
the resulting expressions coincide with the formulas presented in \cite{Pozrikidis1992}.

The axial Green's function has a logarithmic singularity as $\displaystyle \delta r\rightarrow 0$ (away from the axis $\displaystyle \rho =0$), which can be isolated using the asymptotics of elliptic functions at infinity:
\begin{equation*}
K(-\nu )\sim \frac{\ln (4\sqrt{\nu } )}{\sqrt{\nu }} ,\ E(-\nu )\sim \sqrt{\nu } +\frac{1}{2}\ln \nu /\sqrt{\nu } ,
\end{equation*}

\begin{equation*}
G_{\mu \nu } \sim \delta _{\mu \nu }\textcolor[rgb]{0.82,0.01,0.11}{\frac{2}{\rho }}\textcolor[rgb]{0.82,0.01,0.11}{\ln}\textcolor[rgb]{0.82,0.01,0.11}{(4}\textcolor[rgb]{0.82,0.01,0.11}{\frac{2\rho }{\delta r}}\textcolor[rgb]{0.82,0.01,0.11}{)} \ .
\end{equation*}

\end{document}